\documentstyle[12pt]{article}
\newcommand{\tas}{{\rm Tr}({\cal H})_{\rm s}}
\newcommand{\tiso}{{\rm Tr}({\cal H})_{\rm iso}}
\newcommand{\tra}{{\rm Tr}({\cal H})}
\newcommand{\be}{\begin{equation}}
\newcommand{\ee}{\end{equation}}
\newcommand{\sli}{S_{\rm lin}}

\newcommand{\eev}{e_1\vee e_2}
\newcommand{\dem}{d\mu(\zeta)}
\newcommand{\kai}{\kappa\int\limits_Q}
\newcommand{\cala}{\int\limits_{-\infty}^{\infty}}
\begin{document}
\title{Effectively classical quantum states for open systems}
\author{Ph. Blanchard$^{a,}$ and R. Olkiewicz$^b$\\
{\small $^a$ Physics Faculty and BiBoS, University of
Bielefeld, 33615 Bielefeld, Germany}\\
{\small $^b$ Institute
of Theoretical Physics, University of Wroc{\l}aw, 50-204 Wroc{\l}aw, Poland}}

\maketitle
\begin{abstract}
Notions of robust and "classical" states for an
open quantum system are introduced and discussed in the framework of
the isometric-sweeping decomposition of trace class
operators. Using the predictability sieve proposed
by Zurek, ``quasi-classical'' states are defined. A
number of examples illustrating how
the ``quasi-classical'' states correspond to classical points in phase space
connected with the measuring apparatus are presented.
\end{abstract}

\newpage
\noindent
{\bf 1. Introduction}

\vspace{4mm}
Quantum mechanics, whose basic laws were formulated in the twenties, still remains the most fundamental
theory we know. Although, it was originally conceived as a theory of atoms, it has shown a wide range of
applicability, making it more and more evident that the formalism describes some general properties of
Nature. However, despite its successes, there is still no consensus about its interpretation with the main
questions being centered around the quantum measurements. 
Clearly, the existence of classical quantities which would allow to express the measurement
results and explain the classical appearance of the macroscopic world is a fundamental problem in this
matter. The standard explanation which states that, for example, the center-of-mass motion of a macroscopic
object should be described by a narrow wave packet, well localized in both position and momentum, is not
satisfactory. It still remains unanswered why such objects are represented by narrow wave packets, while the
superposition principle allows the emergence of non-classical states as well. Moreover, measurement-like
processes would necessarily produce such non-classical states, as in the infamous example of Schr\"odinger's
cat. A superposition of being dead and alive should produce an entirely new state, in the same sense as
the superposition of $K$ meson and its antiparticle does.

The main reason for this annoying situation seems to be
based on the assumption that it is
possible to isolate systems from their environment.
When we drop it as unjustified and consider quantum
systems as open ones we obtain a new perspective for
the understanding of the emergence of classical
properties within the framework of open systems theory.
This is the basic objective of the program of decoherence
proposed by Zurek [1,2] and further developed in [3,4,5].
For a recent review of the subject and a wide range of references up to
1996 see [6]. Decoherence is a process of continuous measurement-like
interaction between a system and its environment which results in limiting
the validity of the superposition principle in the Hilbert space of the
system. In other words, the environment destroys the vast majority of
superpositions in short time, and, in the case of macroscopic objects,
almost instantaneously. This leads to the appearance of environment-induced
superselection rules, which precludes all but a particular subset of states
from stable existence. On the other hand, it singles out a preferred set of
states which behave in an effectively classical, predictable
manner. Generalizing this notion the predictability sieve was introduced
[7,8] (see also [9]). It is a procedure, which systematically
explores states of an open quantum system in order to arrange them and next
put on a list, starting with the most predictable ones and ending with
those, which are most affected by the environment. Clearly, the states being
on the top of the list can be thought of as "classical" or ``quasi-classical''
ones.

In order to study decoherence, the analysis of the evolution
of the reduced density matrix obtained by
tracing out the environment variables is the most
convenient strategy. If the interaction is such that
the reduced density matrix becomes approximately
diagonal in a particular basis (in the simplest case),
then it is said that an environment-induced
superselection structure has emerged. Generally, the procedure
of tracing out environment variables, being
the composition of a unitary automorphism
with a conditional expectation, leads to a
complicated integro-differential equation for the
reduced statistical operator.
However, for a large class of interesting
physical phenomena we can derive, using certain
limiting procedures, an approximate
Markovian master equation for the reduced density matrix [10]. More
recently, the derivation of the master equation for the reduced density
matrix of a system coupled linearly to an ohmic, subohmic and
supraohmic environment at arbitrary temperature has been obtained in
[11].

Usually, when deriving the master equation for the
reduced density matrices it is assumed that the quantum
system interacts with the environment, which is
another quantum system and hence is also described in terms
of quantum mechanics. However, it should be
pointed out here that such an assumption stemming from the thinking
of quantum mechanics as a universal theory,
can be replaced by a more general one. Sometimes, it is more
useful and natural to treat the external degrees of
freedom as a classical system described by a commutative
algebra of functions. In this approach ( see [12,13] for
discrete classical systems and [14,15] for
continuous ones) the evolution equation for the
classical part is modified by the expectation value of
some quantum observable while, at the same time,
the Schr\"odinger unitary dynamics for the quantum
subsystem is replaced by a dynamical semigroup of
completely positive maps. Therefore, when we allow the
quantum system to interact with its environment, then,
regardless of the nature of this interaction, the
evolution becomes dissipative, given by the Markovian master equation.

The loss of quantum coherence in the Markovian
regime was established in a number of open systems [16,17],
giving a clear evidence of environment-induced superselection
rules. In a recent paper [18] a thorough
mathematical analysis of the superselection structure
associated to an environment-induced semigroup was presented.
It was achieved by the use of the isometric-sweeping
decomposition, which singles out a subspace of density
matrices, on which the semigroup acts in a reversible,
unitary way, and sweeps out the rest of statistical
states. The purpose of this paper is to pursue that
investigation with a particular emphasis on the analysis
of the classicality of states. We put the notion of pointer states,
previously introduced and discussed by Zurek and other authors,
into a general framework, and examine
their properties. It is worth noting that proposed definitions are
expressed solely in terms of the dynamical semigroup and does not refer to
any additional conditions like that one involving the knowledge of the
state of the quantum system before the beginning of interaction.
A number of examples including that of
a quantum stochastic process of Davies, and
illustrating how the ``quasi-classical''
states correspond to classical points in the underlying phase space are
also presented.\\[4mm]
{\bf 2. "Classical" states}

\vspace{4mm}
One can show that purely unitary evolution can never
resolve the apparent conflict between predictions it
implies and perception of the classical reality.
Therefore, in order to explain the appearance of classical
(non-quantum) properties of a quantum system,
we have to open the system and allow it to
interact with the environment. As was mentioned in
Introduction we restrict our considerations to the
Markovian regime and thus assume that the evolution
of the reduced density matrix is given by an
environment-induced semigroup. The concept of the environment-induced
semigroup was introduced in [19],
and the justification of the name was also given there.
They form a subclass of dynamical semigroups
(completely positive, trace preserving and contractive in
the trace norm $\|\cdot\|_1$), which are also
contractive in the operator norm $\|\cdot\|_{\infty}$.
This additional property ensures that both the
linear and statistical entropy of the open quantum system never
decrease in the course of interaction [19].

In this section we search for quantum states which can correspond to the
classical points in phase space
once the interaction with the environment is acknowledged. Our strategy
is as follows. We start with a
quantum system whose evolution is given by an environment-induced
semigroup $T_t$ without asking question
where it comes from. Then, using general principles and properties of
that semigroup we determine sets of
"classical" and ``quasi-classical'' (see the next section) states. If they
are empty and all states
are stable, the system can be
thought of as closed, evolving in a unitary way, and hence the semigroup
may be extended to a one
parameter group of unitary automorphisms. However, if one of them is
non-trivial, then the system is
open and we conclude that the selected states correspond with points in
a classical phase space.
Therefore, these sets contain the information about the type of
interaction with the environment
(measuring apparatus) which led to the appearance of the semigroup $T_t$.

   At first, let us comment on crucial differences
   between states of classical and quantum systems. One of the most
   characteristic features distinguishing
   classical from quantum states is their sensitivity to measurements. In
   classical physics we could perform
   many kinds of measurements which would not disturb the system in an
   essential way. A measurement can
   increase our knowledge of the state of the system but, in principle, it
   has no effect on the system
   itself. By contrast, in quantum mechanics it is impossible to find out
   what the state is without, at
   the same time, changing it in the way determined by the measurement.
   According to the von Neumann
   projection postulate the outcome will be, in general, represented by a
   density matrix. Therefore, as a
   convenient measure of the influence of the environment on the state, we
   take the measure of the loss
   of its purity expressed in terms of the linear entropy
   $\sli(\rho)\,=\,{\rm tr}(\rho\:-\:\rho^2)$
   [5,8].

   Let $\cal S$ denote the set of all states of the quantum system. By a
   state we always mean a pure
   state, whereas for a mixed state we reserve such notions
   like density matrix or statistical state.
   Hence $\cal S$ consists of unit vectors from a Hilbert space $\cal H$
   determined up to the phase factor $[|\psi>]$
   or, in other words, of one-dimensional projectors in $\cal H$. Hence
   $[|\psi>]$ is the abstract class of unit
   vectors with respect to the following equivalence relation:
   $|\psi>\equiv |\psi'>$ if $|\psi>\,=\,e^{i\alpha}
   |\psi'>$ for some $\alpha\in{\bf R}$. Let us notice that the scalar
   product of two distinct states is not well
   defined but its absolute value is. Also the one-dimensional projector
   $|\psi><\psi|$ does not depend on the
   choice of a state vector $|\psi>$.

   In order to define a subset of robust (completely stable)
   states let us notice that any environment-induced semigroup $T_t$
   determines two linear closed and $T_t$-invariant subspaces $\tiso$ and
   $\tas$ in the Banach space of all
   trace class operators $\tra$. The subspace $\tiso$ is called the
   isometric part and $\tas$ the sweeping
   part. For the reader convenience we recall here some basic results of
   this isometric-sweeping decomposition.
   For proofs and a more detailed discussion see [18]. The isometric and
   sweeping subspaces have the
   following properties:\\
   a) $\tiso$ and $\tas$ are $^*$-invariant,\\
   b) $\tiso\bot\tas$ in the following sense:
   $\forall\phi_1\in\tiso\;\forall\phi_2\in\tas$ we have
   ${\rm tr}\phi_1\phi_2\,=\,0$,\\
   c) $\tra\,=\,\tiso\oplus\tas$, $T_t\,=\,T_{1t}\oplus T_{2t}$,\\
   d) $T_{1t}$ is an invertible isometry given by a unitary group, i.e.
   $T_{1t}\phi\,=\,U_t\phi U_t^*$
   for any $\phi\in\tiso$,\\
   e) $T_{2t}$ is sweeping, i.e. $w^*-\lim_{t\to\infty}T_{2t}\phi\,=\,0$
   for any $\phi\in\tas$, where
   $w^*$ denotes the weak$^*$ topology.\\
   Hence, to any environment-induced semigroup corresponds a space
   of statistical states $\tiso$ and associated with it
   an algebra of observables such that the evolution,
   when restricted to these spaces, is given by the
   Schr\"odinger unitary dynamics. In addition, $\tiso$ has the following
   properties:\\
   (i) if $\phi_1,\,\phi_2\in\tiso$, then also
   $\phi_1\cdot\phi_2\in\tiso$,\\
   (ii) if projectors $e,\,f\in\tiso$, then also $e\vee f\in\tiso$, where
   $e\vee f$ denotes a projector
   onto the two-dimensional subspace spanned by the ranges of $e$ and
   $f$.\\
   As a consequence, any one-dimensional projector $e\in\tiso$ remains a
   projector during the evolution,
   and so $\sli(T_te)\,=\,0$ for any $t\geq 0$. Therefore, we define a
   subset ${\cal S}_0$ of
   robust states by
   \be {\cal S}_0\;=\;{\cal S}\cap\tiso\ee
   or, equivalently,
   $${\cal S}_0\;=\;\{e\in{\cal S}:\;\:\sli(T_te)\;=\;\sli(T^*_te)\;=\;0\}$$
   where $T^*_t$ denotes the adjoint semigroup. If $T^*_t$ commutes with
   $T_t$, then
   $${\rm
   tr}(T^*_te)^2\;=\;<T^*_te,\,T^*_te>_{HS}\;=\;<T_te,\,T_te>_{HS}\;=\;
   {\rm tr}(T_te)^2$$
   where $<\cdot,\cdot >_{HS}$ is the scalar product in the Hilbert space
   of Hilbert-Schmidt operators,
   and so the condition $\sli(T_te)\,=\,0$ implies that also
   $\sli(T^*_te)\,=\,0$. Therefore, in such a case for a state
   $e\in{\cal S}$ to be robust it is enough that its linear  entropy does
   not change in the course of evolution. It is worth noting that
   for quantum systems over finite dimensional Hilbert spaces it also turns
   out that the condition
   $\sli(T_te)\,=\,0$ alone is sufficient for state $e$ to be in
   ${\cal S}_0$, see Appendix. If
   ${\cal S}_0=\,{\cal S}$, then the semigroup $T_t$ may be extended to a
   group of unitary automorphisms.

   Obviously, any state from ${\cal S}_0$ will remain pure during the
   evolution and so remain in ${\cal S}_0$.
   Therefore, elements from ${\cal S}_0$ are the most probable candidates
   for ``classical'' states. But
   the unitary evolution and thus perfect predictability alone does not
   suffice to
   accomplish our goal. Another feature distinguishing
   quantum from classical states, namely the validity of the superposition
   principle, has to be taken into
   account. In quantum mechanics it guarantees that any superposition of
   two distinct, and not necessarily
   orthogonal, states is again a legitimate quantum state. It means that
   for any
   pair of different one-dimensional projectors $e_1$ end $e_2$ we can
   associate a set of one-dimensional
   projectors $e$ given by $e\cdot(\eev)\,=\,e$. Equivalently, we may
   write that
   $$e\;=\;\frac{(z_1|\psi_1>\:+\:z_2|\psi_2>)(z^*_1<\psi_1|\:+\:z^*_2
   <\psi_2|)}{\|z_1|\psi_1>\:+\:z_2|\psi_2>\|^2}$$
   where $|\psi_1><\psi_1|\,=\,e_1$, $|\psi_2><\psi_2|\,=\,e_2$ and
   $z_1,z_2$ are complex numbers.
   By contrast, classical states do not combine into another state. The
   only situation when their combination
   can be considered is inevitably tied to probability distributions on the
   phase space.
   Therefore, it is natural to assume that any non-trivial superposition of
   "classical" states cannot be robust.\\
   {\bf Definition 2.1}. {\it A state $e\in{\cal S}$ is called
   "classical" if
   $e\in{\cal S}_0$ and for any $f\in{\cal S}_0$, $f\neq e$,
   $S(e,\,f)\cap{\cal
   S}_0\,=\,\emptyset$, where $S(e,\,f)$ denotes the collection of all
   states being non-trivial
   superpositions of $e$ and} $f$.
   {\it The collection of all "classical" states we denote by}
   ${\cal S}_c$.\\
   Hence, although
   "classical" states remain pure during the evolution, any of their
   superpositions deteriorates into a mixture. Under a mild, technical
   assumption namely that $T_t$ admits a holomorphic extension to a sector
   $\sum_{\epsilon}\,=\,\{z:\;{\rm Re}z>0,\,|{\rm arg}z|<\epsilon\}$, for
   some $\epsilon>0$, the loss
   of the purity of their superpositions happens instantaneously.

   We are now in position to describe the structure of set ${\cal S}_c$.\\
   {\bf Theorem 2.2}. If ${\cal S}_c\neq\emptyset$, then it consists of a
   family, possibly finite, of pairwise orthogonal states
   $\{e_1,\,e_2,...\}$
   such that $T_te_i\,=\,e_i$ for all $t\geq 0$ and any index $i$.\\
   Proof: Let $e\in{\cal S}_c$. We show that $e$ is
   orthogonal to any state $f\in{\cal S}_0$, $f\neq e$. Suppose, on the
   contrary, that $e\cdot f\neq 0$. Then, because $e\vee f\in\tiso$, the
   state
   $e'=\,e\vee f\,-\,e$ also belongs to $\tiso$ and is orthogonal to $e$.
   All of
   these states can be considered as acting on a two-dimensional Hilbert
   space,
   the range of $e\vee f$. Choosing an appropriate coordinate system we
   represent them by
   $$e\;=\;\frac{1}{2}(I\:+\:\vec{n}_1\cdot\vec{\sigma}),\;f\;=\;\frac{1
   }{2}(I\:+\:\vec{n}_2\cdot\vec{\sigma})$$
   with $\vec{n}_1\,=\,(0,\,0,\,1)$ and
   $\vec{n}_2\,=\,(\cos\theta,\,0,\,\sin\theta)$, $\theta\in
   [0,\,\pi/2)$. Because the hermitian matrix $i[e,\,f]\in\tiso$ is
   non-zero so
   its spectral projectors $\frac{1}{2}(I\:\pm\:\vec{m}\cdot\vec{\sigma})$,
   where $\vec{m}\,=\,(0,\,1,\,0)$, also belong to $\tiso$. On the other
   hand, they are superpositions of $e$ and $e'$. Therefore,
   $S(e,\,e')\cap{\cal
   S}_0\neq\emptyset$, what contradicts the assumption that $e$ is
   "classical". Hence $e\bot f$ for any $f\in{\cal S}_0$, $f\neq e$ and
   so
   ${\cal S}_c\,=\,\{e_1,\,e_2,...\}$ with
   $e_i\cdot e_j\,=\,\delta_{ij}e_i$. Finally, we show that
   $T_te\,=\,e$ for all
   $t$. If not so, then for any $\epsilon>0$ we find an instant $s$ such
   that
   $T_se\neq e$ and $\|T_se\:-\:e\|_1<\epsilon$.
   However, by the above argument, $T_se$ is orthogonal
   to $e$, so $\|T_se\:-\:e\|_1\,=\,2$, the contradiction. $\Box$\\
   Therefore, it turned out that "classical" states, which are defined in
   a general way,
   form so-called pointer basis being introduced so far only on the
   operational level.
   Let us recall that pointer basis
   arises in a specific situation when before the measurement the quantum
   system was in an eigenstate
   of the measured observable. Such states are completely predictable since
   they do not evolve at all.
   By Theorem 2.2, they always correspond to points in a discrete classical
   phase space.

   It is also clear that a unitary evolution
   $T_t\,=\,e^{-itH}\cdot e^{itH}$ with $H\,=\,H^*$, does not lead to
   the appearance of
   "classical" states at all. Although ${\cal S}_0=\,{\cal S}$ in this
   case, ${\cal S}_c=\,\emptyset$
   since any superposition of robust states is again robust.
   It is worth noting that, in general, even if "classical" states exist,
   they may form an incomplete set of one-dimensional projectors.\\[4mm]
   {\bf 3. ``Quasi-classical'' states}

   \vspace{4mm}
   In this section we continue the investigation of states of an open
   quantum system which offer optimal
   predictability of their own future values. In the case when
   ``classical''
   states exist, they are the best candidates for states corresponding to
   classical points of phase space. If
   they are absent, it is natural to consider the states which are least
   affected by the interaction with the
   environment, that is, which are least prone to deteriorate into
   mixtures. Since linear entropy is a convenient
   measure of the loss of purity, we take its increase for initial states
   as a basic criterion. For a more complete analysis one
   should search for states which minimize the linear entropy over some
   finite period of time characteristic for
   the evolution of the system. To start with we define a quadratic form on
   the Hilbert space HS$({\cal H})$ of
   Hilbert-Schmidt operators
   $$B(\phi)\;=\;-<\phi,\,L(\phi)>_{HS}$$
   where $\phi\in D(L)\subset\tra$ and $L$ denotes the generator of
   semigroup $T_t$. Since $\tra$ is dense in
   HS$({\cal H})$ so $B$ is densely defined. The closure of its symmetric
   part we denote by $\lambda$. By the
   Hille-Yosida theorem, the Lumer-Philips form, $\lambda$ is positive
   definite. It is clear that
   \be \lambda(\phi)\;=\;\frac{1}{2}\frac{d}{dt}\sli(T_t\phi)|_{t=0}\ee
   whenever the corresponding derivative exists. Let
   \be {\cal S}(a)\;=\;\{e\in{\cal S}\cap
   D(\lambda):\;\:\lambda(e)\:=\:a\}\ee
   for $a>0$, and put $a_0\,=\,\inf\{a:\;{\cal S}(a)\neq\emptyset\}$.
   Guided by the previous considerations we define
   the set ${\cal S}_s$ of most stable states by ${\cal S}_s\,=\,{\cal
   S}(a_0)$, if it is non-empty. When ${\cal S}
   (a_0)=\emptyset$, then, in general, ${\cal
   S}_s\subset\bigcup_{a<a_0+\epsilon}{\cal S}(a)$. The choice of
   $\epsilon$ is somewhat arbitrary as it serves as the border between the
   preferred ``quasi-classical'' states and the
   "non-classical" remainder. In this case, as was mentioned above, a
   further analysis examining the behavior
   of $T_te$ also for $t>0$ may be inevitable in order to select the set
   ${\cal S}_s$.

   By combining predictability with the previously exploited principle expressing
   the fact that any superposition of two distinct
   preferred states cannot belong to the same class of stability we obtain
   the following.\\
   {\bf Definition 3.1}. {\it A state $e$ is called
   ``quasi-classical'' if $e\in{\cal S}_s$ and for any} $f\in{\cal S}_s$,
   $f\neq e$, $S(e,\,f)\cap{\cal S}_s\,=\,\emptyset$.
   {\it The space of ``quasi-classical'' states will be denoted by} ${\cal S}_{qc}$.\\
   It should be pointed out that ``quasi-classical''
   states can form an overcomplete set in contrast to the "classical"
   states. On the other hand
   they may not exist at all. A simple example illustrating such a case is
   given by a dynamical semigroup on
   $2\times 2$ complex matrices with the following generator:
   $$ L(\rho)\;=\;-i[H,\,\rho]\;+\;({\rm tr}\rho)I\;-\;2\rho$$
   Then ${\cal S}(a)\,=\,\emptyset$ if $a\neq 1$ and ${\cal
   S}(a)\,=\,{\cal S}$ for $a\,=\,1$. Hence ${\cal S}_s$
   consists of all states and so any superposition of its two states again
   belongs to ${\cal S}_s$. Therefore
   ${\cal S}_{qc}\,=\,\emptyset$. In this case all states deteriorate into a
   completely mixed state in a uniform way.\\[4mm]
{\bf 4. Examples}

\vspace{4mm}
Having discussed theoretical properties of
"classical" and ``quasi-classical'' states, let us now consider some physical examples.\\[3mm]
{\it 4.1. Pointer states}

\vspace{3mm}
Pointer states have been thoroughly discussed, see [6] and references
therein. They arise, for example, when the dynamical generator for
the reduced density matrix is given by (see [12,13])
$$L(\rho)\;=\;-i[H,\,\rho]\;+\;\sum_iP_i\rho P_i\;-\;\frac{1}{2}
\{P,\,\rho\}$$
where $P_i$ are one-dimensional orthogonal projectors, $P\,=\,\sum_i
P_i$, and Hamiltonian $H$ commutes with all $P_i$. Then ${\cal S}_c=\,
\{P_i\}$.\\[3mm]
{\it 4.2. Quantum Brownian motion}

\vspace{3mm}
For quantum Brownian motion
\be\dot{\rho}\;=\;-i[H,\,\rho]\;-\;D[x,\,[x,\,\rho]]\ee
which leads to an environment-induced semigroup,
the rate of change of linear entropy for a state
$e\,=\,|\psi><\psi|$ is given by
$$\frac{d}{dt}\sli(T_te)|_{t=0}\;=\;4D(<x^2>\:-\:<x>^2)$$
where $<x>\,=\,<\psi|x|\psi>$ and $<x^2>\,=\,<\psi|x^2|\psi>$.
Hence $\lambda(e)$ is proportional to the
dispersion in position of $|\psi>$ and so
${\cal S}(a)\neq\emptyset$ for any $a>0$. However, ${\cal S}(a=0)\,
=\,\emptyset$. A more detailed analysis shows
that the space of the most stable states ${\cal S}_s$ consists
of coherent states of the quantum harmonic oscillator [7].
Because these states are represented by [20]
$$|\alpha>\;=\;\exp(-\frac{1}{2}|\alpha|^2)\sum\limits_{n=0}^{\infty}
\frac{\alpha^n}{(n!)^{1/2}}|n>$$
where $\alpha$ is a complex number and $|n>$
denotes the energy eigenstate, hence for any two distinct states
$|\alpha>$ and $|\beta>$, $\alpha\neq\beta$,
none of their superpositions belongs to ${\cal S}_s$. Therefore,
${\cal S}_s\cap S(|\alpha><\alpha|,\,|\beta><\beta|)\,=\,\emptyset$,
and so the coherent states are ``quasi-classical''. It is worth noting that
coherent states of a harmonic oscillator coupled with up and down spins are
also selected as the preferred states (``quasi-classical'' in our terminology)
in a model of the joint system of a spin-$\frac{1}{2}$ particle and a harmonic oscillator
interacting with a zero-temperature bath of harmonic oscillators [21].\\[3mm] 
{\it 4.3. GRW spontaneous localization}

\vspace{3mm}
Let us now examine the behavior of pure states
$e_{\psi}=\,|\psi><\psi|$ for a semigroup
given by the master equation of the type  discussed by
Ghirardi, Rimini and Weber [22] (see also [23])
\be\dot{\rho}\;=\;-i[H,\,\rho]\;+\;\cala da\,G_a\rho G_a\;-\;\kappa\rho\ee
where $G_a$ is an operator of multiplication by a Gaussian function
$$g_a(x)\;=\;\kappa^{1/2}(\frac{2\alpha}{\pi})^{1/4}e^{-\alpha(x\:-\:a)^2}$$
that is $G_a\psi(x)\,=\,g_a(x)\psi(x)$. Clearly, the above
master equation leads to a dynamical semigroup
which is also contractive in the operator norm. Hence $\lambda$ is well
defined and$$\lambda(e_{\psi})\;=\;\kappa\;-\;\cala da({\rm
tr}\,G_ae_{\psi})^2$$$$=\;\kappa\;-\;\cala\cala
dxdy|\psi(x)|^2|\psi(y)|^2\cala
da\,g_a(x)g_a(y)$$$$=\;\kappa[1\:-\:\cala\cala
dxdy|\psi(x)|^2|\psi(y)|^2e^{-\alpha(x\:-\:y)^2/2}]$$Therefore, for any
$e_{\psi}$, $0<\,\lambda(e_{\psi})<\kappa$ and for any $0<a<\kappa$, ${\cal
S}(a)\neq\emptyset$. Clearly, the most stable states are those of Dirac's
delta type, whereas states which are uniformly distributed over large
intervals are strongly affected by the interaction.\\[3mm]
{\it 4.4. Quantum stochastic process}

\vspace{3mm}
This example shows that coherent states can be also
selected as the ``quasi-classical'' states for quantum
stochastic processes introduced by Davies [24].
Quantum stochastic processes
were introduced to describe rigorously certain continuous
measurement processes. They can be constructed from
two infinitesimal generators. The first is the
generator $Z$ of a strongly continuous semigroup
on a Hilbert space $\cal H$, and the second is a stochastic
kernel $J$, describing how the measuring apparatus
interacts with the system. Let us recall that a stochastic
kernel is a measure defined on the $\sigma$-algebra
of Borel sets in some locally compact space and with
values in the space of bounded positive linear
operators on $\tra$. In this example we take Poincar\'e disc
$D\,=\,\{\zeta\in{\bf C}:\;|\zeta|<1\}$ as the
underlying topological space, and define
$$Z\;=\;iH\;-\;\frac{\kappa}{2}{\bf 1}$$
where $H$ is the Hamiltonian of the system,
$\kappa>0$ is the coupling constant. For $E\subset D$ and $\rho
\in\tra$ the stochastic kernel is defined by
$${\rm tr}[J(E,\,\rho)A]\;=\;\kappa\int\limits_E\dem
{\rm tr}(e_{\zeta}\rho e_{\zeta}A)$$
where $A$ is a bounded linear operator on $\cal H$,
$e_{\zeta}\,=\,|\zeta><\zeta|$ with $|\zeta>$ being a
SU(1,1) coherent state, i.e. a holomorphic function on $D$ [25]
$$|\zeta>(z)\,=\,(1\:-\:|\zeta|^2)(1\:-\:z\zeta)^{-2}$$
and
$$\dem\;=\;\frac{1}{\pi}\frac{d\zeta d\bar{\zeta}}{(1\:-\:|\zeta|^2)^2}$$
is a SU(1,1) invariant measure on $D$. In order to define a
quantum stochastic process $Z$ and $J$ have to
satisfy the following relation
$${\rm tr}[J(D,\,e_{\psi})]\;=\;-2{\rm Re}<\psi|Z|\psi>$$
$e_{\psi}=\,|\psi><\psi|$, for all normalized vectors
$\psi\in D(Z)$. It is straightforward to check that
$${\rm tr}[J(D,\,e_{\psi})]\;=\;\kai\dem {\rm tr}
(e_{\zeta}e_{\psi}e_{\zeta})\;=\;\kappa\;=\;-2{\rm Re}<\psi|Z|\psi>$$
The strongly continuous semigroup $T_t$ associated
with the process is given by
$$T_t(\rho)\;=\;\exp(tZ^*)\rho\exp(tZ)\;+\;tJ(D,\,\rho)\;+\;o(t)$$
and so its generator reads
\be L(\rho)\;=\;-i[H,\,\rho]\;+\;\kai\dem e_{\zeta}\rho
e_{\zeta}\;-\;\kappa\rho\ee
Obviously, it generates an environment-induced semigroup. It
is worth noting that the integral formula above is a straightforward
generalization of the von Neumann projection postulate to the case in which the
family of states is overcomplete. Such a generator was thoroughly
discussed in [26].

We now search for the most stable states
with respect to semigroup $T_t$. Let us first note that no state is
stable since, by Lemma 3.2 in [26], $T_t$ is strictly positive, that is
$T_te$ is a faithful density matrix for every $t>0$ and any $e\in{\cal S}$.
Hence ${\cal S}_0\,=\,\emptyset$ and so, in this case, there are no
"classical" states at all. However, the quadratic form $\lambda$ is bounded
and allows to classify all states in the following way.\\
{\bf Proposition 4.1}.
$\frac{2}{3}\kappa\leq\lambda(e)<\kappa$ for every $e\in{\cal S}$, and
${\cal S}(a)\neq\emptyset$ if $\frac{2}{3}\kappa\leq a<\kappa$. For
$a_0\,=\,\frac{2}{3}\kappa$, ${\cal S}(a_0)$ consists exactly of the coherent
states $e_{\zeta}$, $|\zeta|<1$.\\
Proof: Let $e\in{\cal S}$. Then, by definition,
$$\lambda(e)\;=\;\kappa[1\:-\:\int\limits_Qd\mu(\zeta)({\rm
tr}\,ee_{\zeta})^2]$$
Suppose $e_n=\,|n><n|$, where
$|n>(z)\,=\,\sqrt{n\,+\,1}z^n$, $n\in{\bf N}\cup\{0\}$, is an
orthonormal basis in $\cal H$. It was shown in [15] that
$$\int\limits_Qd\mu(\zeta)({\rm
tr}\,e_ne_{\zeta})^2\;=\;\frac{n\:+\:1}{(2n\:+\:1)(2n\:+\:3)}$$
Therefore,
$$0\:<\:\int\limits_Qd\mu(\zeta)({\rm
tr}\,ee_{\zeta})^2\:\leq\:\frac{1}{3}$$
for any $e$, and so
$\frac{2}{3}\kappa\leq\lambda(e)<\,\kappa$. Finally, notice that $\lambda$
is SU(1,1) invariant. Hence
$\lambda(e_0)\,=\,\lambda(\pi(g)e_0\pi(g)^*)\,=\,\frac{2}{3}\kappa$ for
any $g\in$SU(1,1). However, the set $\{\pi(g)e_0\pi(g)^*\}$ coincides with
the set of coherent states $e_{\zeta}$, $|\zeta|<1$. $\Box$\\
Hence, by
definition, ${\cal S}_s\,=\,{\cal S}(a_0)$. Not surprisingly, the most
stable states are the coherent ones. Next we show that they are
``quasi-classical''. Because any $|\zeta>$ has the following
representation
$$|\zeta>\;=\;(1\:-\:|\zeta|^2)\sum\limits_{n=0}^{\infty}
\sqrt{n\:+\:1}\zeta^n|n>$$
it follows that
any superposition of two distinct coherent states $|\zeta>$ and $|\zeta'>$
is not coherent, and thus
${\cal S}_s\cap S(e_{\zeta},\,e_{\zeta'})\,=\,\emptyset$.
Hence the coherent states are ``quasi-classical'' and make up an analog of the pointer basis. They
correspond to points in Poincar\'e disc, which is the underlying space of
the stochastic kernel representing the measuring apparatus.

In this section we have presented different types of examples which demonstrate how robust states
can be selected and how non-stable states decohere to mixtures. Moreover, the rate of their
deterioration to density matrices was examined. Such an analysis can be useful, for example, in
quantum computing, where it is essential to control the process of decoherence in order to
allow quantum bits to compute in parallel.\\[4mm]
{\bf Acknowledgements}

\vspace{4mm}
One of us (R.O.) would like to thank the A. von Humboldt Foundation
for the financial support.\\[4mm]
{\bf Appendix}

\vspace{4mm}
Suppose dim${\cal H}<\infty$ and let $\sli(T_te)\,=\,0$ for some
   $e\in{\cal S}$. Then
   $e\in{\cal S}_0$.\\
   Proof: Let $P$ be the projection onto $\tiso$ along $\tas$.
   Because $P$ extends to an orthogonal projection in the Hilbert space
   of Hilbert-Schmidt operators on $\cal H$ so
   $$\|T_te\|_2^2\;=\;\|P(T_te)\|_2^2\;+\;\|(id\:-\:P)T_te\|_2^2$$
   where $\|\cdot\|_2$ denotes the Hilbert-Schmidt norm. Since $P$
   commutes with $T_t$ and $T_te\in{\cal S}$, we obtain that
   $$1\;=\;\|T_t(Pe)\|_2^2\;+\;\|(id\:-\:P)T_te\|_2^2$$
   Because dim${\cal H}<\infty$ so $T_t$ is relatively compact in the
   strong operator topology
   and so, by Theorem 24 in [18],
   $\lim_{t\to\infty}\|T_t\phi\:-\:P(T_t\phi)\|_1\,=\,0$ for any
   trace class operator $\phi$. Since $\|\cdot\|_2\leq\,\|\cdot\|_1$ we
   obtain that
   $\lim_{t\to\infty}\|T_t(Pe)\|_2=\,1$. However, $T_t$ is also
   contractive in the norm $\|\cdot\|_2$,
   so $\|Pe\|_2=\,1$ and hence $Pe\,=\,e$. It means that $e\in\tiso$,
   and thus $e\in{\cal S}_0$. $\Box$\\[4mm]
{\bf References}\\[4mm]
$[1]$ W.H. Zurek, Phys. Rev. D 24 (1981) 1516.\\
$[2]$ W.H. Zurek, Phys. Rev. D 26 (1982) 1862.\\
$[3]$ E. Joos and H.D. Zeh, Z. Phys. B 59 (1985) 223.\\
$[4]$ J.P. Paz and W.H. Zurek, Phys. Rev. D 48 (1993) 2728.\\
$[5]$ E. Joos, Decoherence through interaction with the environment, in:
D. Giulini et al. (Eds.), Decoherence and the Appearance of a Classical
World in Quantum Theory, Springer, Berlin, 1996.\\
$[6]$ D. Giulini et al. (Eds.),
Decoherence and the Appearance of a Classical
World in Quantum Theory, Springer, Berlin, 1996.

Ph. Blanchard et al. (Eds.), Decoherence: Theoretical, Experimental and
Conceptual Problems, Lect. Notes Phys. 538, Springer, Berlin, 2000\\
$[7]$ W.H. Zurek, S. Habib and J.P. Paz, Phys. Rev. Lett. 70 (1993) 1187.\\
$[8]$ W.H. Zurek, Progr. Theor. Phys. 89 (1993) 281.\\
$[9]$ M.R. Gallis, Phys. Rev. A 53 (1996) 655.\\
$[10]$ R. Alicki and K. Lendi, Quantum Dynamical Semigroups and Applications,
Lect. Notes Phys. 286, 1987.\\
$[11]$ B.L. Hu, J.P. Paz and Y. Zhang, Phys. Rev. D 45 (1992) 2843.\\
$[12]$ Ph. Blanchard and A. Jadczyk, Phys. Lett. A 175 (1993) 157.\\
$[13]$ Ph. Blanchard and A. Jadczyk, Phys. Lett. A 183 (1993) 272.\\
$[14]$ R. Olkiewicz, Rev. Math. Phys. 9 (1997) 719.\\
$[15]$ R. Olkiewicz, J. Math. Phys. 40 (1999) 1300.\\
$[16]$ W.G. Unruh and W.H. Zurek, Phys. Rev. D 40 (1989) 1071.\\
$[17]$ J. Twamley, Phys. Rev. D 48 (1993) 5730.\\
$[18]$ R. Olkiewicz, Commun. Math. Phys. 208 (1999) 245.\\
$[19]$ R. Olkiewicz, Structure of the algebra of effective observables
in quantum mechanics, quanth-ph/0003032.\\
$[20]$ P. Carruthers and M.M. Nieto, Amer. J. Phys. 33 (1965) 537.\\
$[21]$ A. Venugopalan, Phys. Rev. A 61 (2000) 012102.\\
$[22]$ G.C. Ghirardi, A. Rimini and T. Weber, An attempt at a unified
description of microscopic and macroscopic systems, in: V. Gorini and
A. Frigerio (Eds.), Fundamental Aspects of Quantum Theory, NATO ASI Series
B 144, Plenum Press, New York, 1986.\\
$[23]$ A. Jadczyk, Progr. Theor. Phys. 93 (1995) 631.\\
$[24]$ E.B. Davies, Commun. Math. Phys. 15 (1969) 277.\\
$[25]$ A.M. Perelomov, Commun. Math. Phys. 26 (1972) 222.\\
$[26]$ Ph. Blanchard and R. Olkiewicz, J. Stat. Phys. 94 (1999) 933. 
\end{document}